\documentclass{iau}

\usepackage{amsmath}
\usepackage{graphicx}
\usepackage{multirow}
\usepackage{natbib}

\newcommand {\apj} {ApJ}
\newcommand {\apjs} {ApJS}
\newcommand {\apjl} {ApJ}
\newcommand {\aj} {AJ}
\newcommand {\aap} {A\&A}

\newcommand {\nat} {Nature}

\newcommand {\araa} {ARAA}
\newcommand {\pasp} {PASP}
\newcommand {\physrep} {Phys. Rep.}

\begin{document}

\lefttitle{E.F. van Dishoeck \& MINDS team}
\righttitle{Probing the gas that builds planets with JWST}

\jnlPage{1}{7}
\jnlDoiYr{2024}
\doival{10.1017/xxxxx}

\aopheadtitle{Proceedings of IAU Symposium}
\editors{Z.\ Benkhaldoun, A. Szentgyorgyi, Y. Moulane, eds.}

\title{Probing the gas that builds planets:\\ Results from the JWST MINDS
  program}

\author{Ewine F.\ van Dishoeck and the MINDS team}
\affiliation{Leiden Observatory, Leiden University, the Netherlands}

\begin{abstract}
  Infrared observations with JWST open up a new window into the
  chemical composition of the gas in the inner disk ($<$few au) where
  planets are built.  Results from the MIRI GTO program MINDS are
  presented for several disks around T Tauri and lower-mass stars.
  A large diversity in spectra is found. Some disks are
  very rich in H$_2$O lines whereas other disks show prominent
  CO$_2$. The spectra of disks around very low-mass stars ($<$0.3
  M$_\odot$, late-M type stars like Trappist-1) are dominated by
  C$_2$H$_2$ and other hydrocarbon features including those of
  benzene, suggesting volatile C/O$>$1. Together these data point to a
  rich chemistry in the inner regions that is linked to the physical
  structure of these disks (e.g., dust traps) and that may be affected
  by processes such as radial drift of icy pebbles from the outer to
  the inner disk.
\end{abstract}

\begin{keywords}
protoplanetary disks, planet formation, molecules, infrared spectroscopy
\end{keywords}

\maketitle

\section{Introduction}

Disks around young stars are the birthplaces of planets, but the kind
of planets that are formed and their composition varies significantly
from star to star and is largely determined by the structure and
evolution of their parent disks \citep{Oberg21}. Over the past decade,
new observational facilities at millimeter and infrared wavelengths
have allowed to study the chemical composition of disks with
unprecedented detail. At millimeter wavelengths, the pure rotational
transitions of molecules take place, and the Atacama Large
Millimeter/submillimeter Array (ALMA) has been key to image a growing
number of species \citep{Oberg23}. Both simple and more complex
molecules \citep{Booth24IRS48} have been probed down to abundances as
low as $10^{-12}$ with respected to hydrogen and with fully resolved
line profiles ($R=\lambda/\Delta \lambda = 10^7$). However, ALMA is
limited to the cold outer regions of disks, typically $> 10$ au.

Infrared observations probe the vibrational transitions of molecules,
and, in a few cases (H$_2$O and OH), also very high-lying pure
rotational transitions. The {\it James Webb Space Telescope} (JWST) provides
much higher sensitivity and spectral resolution ($R\approx 3000$) over
the 1--28 $\mu$m range than previous space missions and uniquely
probes the warm inner part of disks ($<$few au) where terrestrial
planets are thought to form. At infrared wavelengths, molecules
without a permanent dipole moment can be observed; this includes H$_2$
as well as carbon-bearing species (e.g., CH$_4$, C$_2$H$_2$, CO$_2$,
...)  down to abundances of $\sim 10^{-8}$. Also Polycyclic Aromatic
Hydrocarbons (PAHs), ices, silicates and other solid-state species are
uniquely probed with infrared data \citep{vanDishoeck04}. Thus, JWST
and ALMA strongly complement each other in studies of the chemistry of
planet-forming disks.

The potential of infrared observations for chemical studies of disks
became apparent from pioneering ground-based observations
\cite[see][for recent summaries]{Brown13,Banzatti22,Banzatti23} as
well as initial space-based data from the {\it Infrared Space
  Observatory} \cite[e.g.,][]{Waelkens96} and the {\it Spitzer Space
  Telescope} \cite[see][for review]{Pontoppidan14}.  This brief paper
summarizes results obtained with the Mid InfraRed Instrument (MIRI) on
JWST \citep{Wright23} in the context of the Mid INfrared Disk Survey
(MINDS) guaranteed time program (PI: Th. Henning, co-PI:
I. Kamp). This 120 hr program observes $\sim$30 disks around solar
mass T Tauri stars, $\sim$10 disks around very low-mass stars and
brown dwarfs and a few Herbig and debris disks with the Medium
Resolution Spectrometer (MRS) at 5--28 $\mu$m
\citep{Kamp23}. Comparison with young disks in the earlier
protostellar stages is done in the JWST Observations of Young
protoStars (JOYS) MIRI-MRS GTO program (PI: E.F. van Dishoeck, co-PI:
H. Beuther) \citep{vanDishoeck23,vanGelder24b}. There is also a
rapidly growing JWST-MIRI data set on protoplanetary disks based on GO
programs, with an additional $\sim$150 disks observed in Cycles 1--3
by other teams, most of them part of the JDISCS collaboration
\citep{Banzatti23JWST,Pontoppidan24}.

The goals of the MINDS program are to use JWST to (1) investigate the
chemical inventory of the terrestrial planet forming zone, (2) 
follow the gas evolution into the disk dispersal stage, and (3)
study the structure of protoplanetary and debris disks in the thermal
mid-IR. The program builds a bridge between the chemical inventory of
planet-forming disks and the properties of exoplanets.  Here only some
highlights related to the first goal will be presented. See
\cite{Henning24} and {\tt minds.cab.inta-csic.es} for an overview and
early summary.

\section{What sets inner disk abundances?}
\label{sect:scenarios}

Three main scenarios affect the inner disk chemistry: (1)
high-temperature gas-phase chemistry; (2) drifting icy pebbles from
the outer disk followed by sublimation once they cross their
snowlines; and (3) dust traps locking up icy pebbles in the outer
disk. In addition, time-variable accretion and ejection processes
(e.g., winds) play a role in all three cases.

Regarding scenario (1): gas temperatures in the inner disk are high,
ranging from 200 K to $>$1500 K, even exceeding the dust sublimation
temperature \citep{Najita03,Woitke18}. UV radiation from the young
stars is also enhanced by several orders of magnitude,
photodissociating and photoionizing molecules and atoms. Thanks to the
high temperatures, reaction barriers can readily be overcome and drive
a rapid chemistry that is different from that taking place in the cold
outer part \citep{Walsh15}. One important case is the balance between
H$_2$O and CO$_2$ through the intermediate OH: at 100--250 K, OH
reacts primarily with CO to form CO$_2$, but at higher temperatures it
reacts with H$_2$ to form H$_2$O \citep{vanDishoeck23}. Because of
continuum optical depth at mid-IR wavelengths, JWST observations
are limited to the upper layers of the disk
\cite[e.g.,][]{Bruderer15}, although grain growth and settling to the
midplane make it possible to look deeper into the disk. Indeed,
effective gas/(small) dust ratios in the infrared emitting layers are
typically 10000 rather than 100 \citep{Meijerink09}.

The C/O ratio is thought to be an important diagnostic for linking
exoplanets with their formation location \citep{Oberg11}. In the inner
disk, dust temperatures are high enough that all ices have sublimated
so the C/O ratio in the gas should be close to the stellar value: the
chemistry reshuffles the individual molecular abundances but does not
affect the overall C/O.

\begin{figure}[t]
\begin{center}
  \includegraphics[width=7cm]{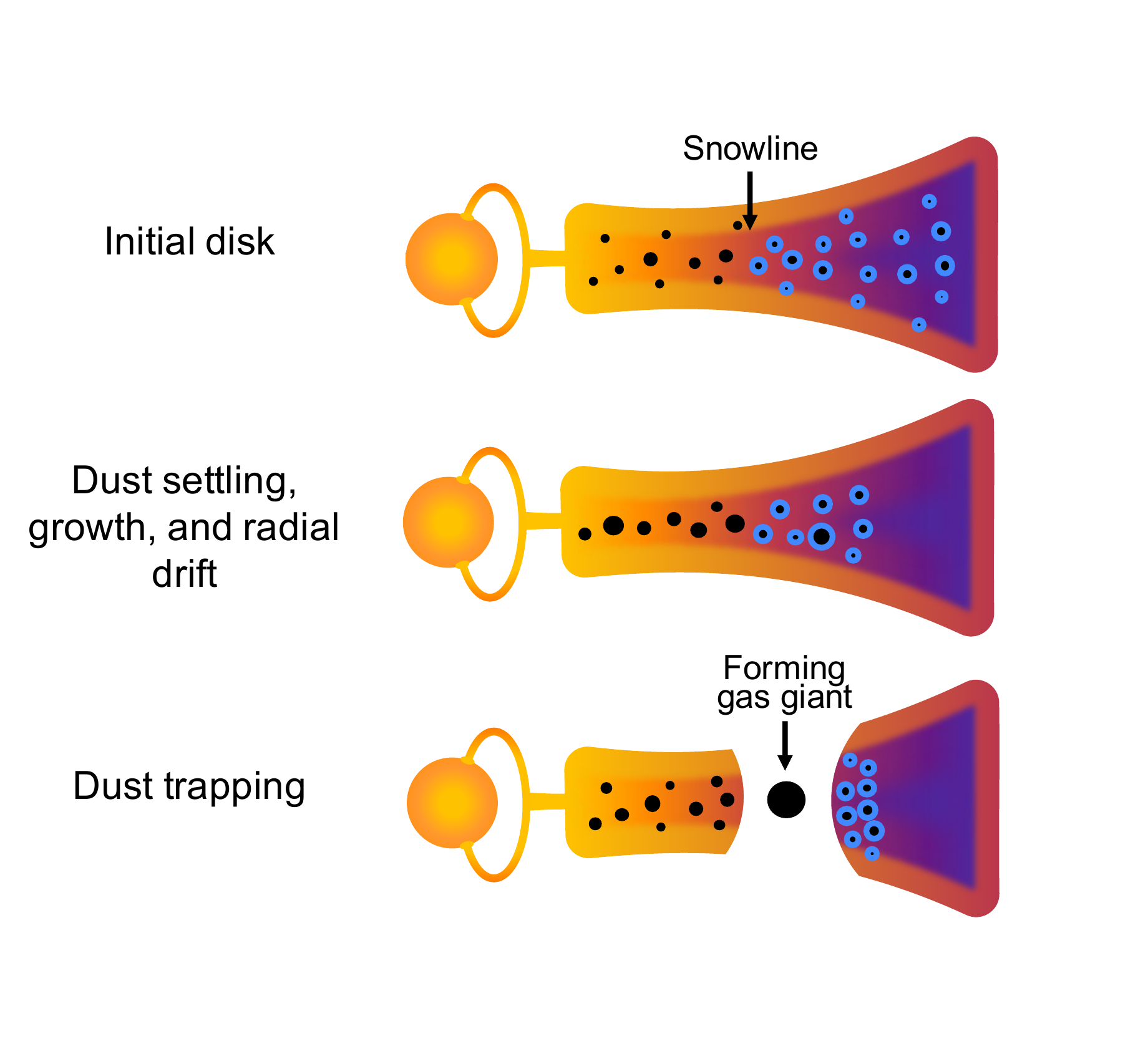}
\end{center}
\vspace{-1.3cm}
\caption{Cartoon illustrating the three scenarios of a full initial
  disk (top), a disk with significant grain growth, settling and
  radial drift of icy pebbles (middle) and a disk with a gap,
  locking up icy pebbles in a dust trap. Figure by Sierra Grant.}
  \label{Fig2}
\end{figure}

Scenarios (2) and (3) alter the C/O (and C/H and O/H) abundances
ratios in the inner region (see Figure~\ref{Fig2}).  In scenario (2),
water (and more generally oxygen, O/H) is enhanced by the drifting of
water-rich icy pebbles from the outer to the inner disk. Once icy
grains cross the disk mid-plane snowline around $\sim$ 160 K, the
water ice sublimates into the gas and becomes visible to JWST if mixed
to high enough altitudes. Enhancements in O/H can be more than an
order of magnitude compared to the interstellar value
\cite[e.g.,][]{Bosman18,Kalyaan21,Mah23}. Indeed, several JWST spectra
show evidence for enhanced ``cold'' ($\sim$200 K) water at the water
snowline \citep{Banzatti23JWST,Banzatti24,Temmink24H2O}. This
enhancement is expected to be particularly strong for disks that show
compact dust millimeter continuum emission with ALMA and large ratios
of their gas/dust radii, indicative of pebbles that have drifted
inwards \citep{Banzatti20,Trapman19}. Similarly, CO and
carbon-containing icy grains can drift in and enhance C/H
\citep[e.g.,][]{Zhang20}.

The opposite case holds for disks with gaps and cavities that show
prominent dust rings (``substructures'') with ALMA
\citep{Andrews20}. In scenario (3), these rings can trap the icy dust
grains and prevent them from moving inwards, thereby lowering the
inner disk O/H and C/H ratios significantly. Much depends on the timing of the
formation of these dust traps \citep{Mah24,Sellek24}: if they form
early ($<$1 Myr), the initial inner disk water rapidly drains onto the
star and the disk becomes very dry.

By comparing inner disk abundances found with JWST with disk
structures imaged by ALMA, scenarios (2) and (3) can be tested. One
caveat is that ALMA can only image cavities at best down to a few au
for nearby disks. This is also the region where the millimeter
continuum becomes optically thick making it more difficult to see
small gaps.  The structure of disks inside a few au is therefore still
largely unknown territory, although near-IR interferometry and
spectrally resolved CO data are providing some constraints.

\section{Results}

\subsection{T Tauri disks}

\begin{figure}[t]
\includegraphics[width=13cm]{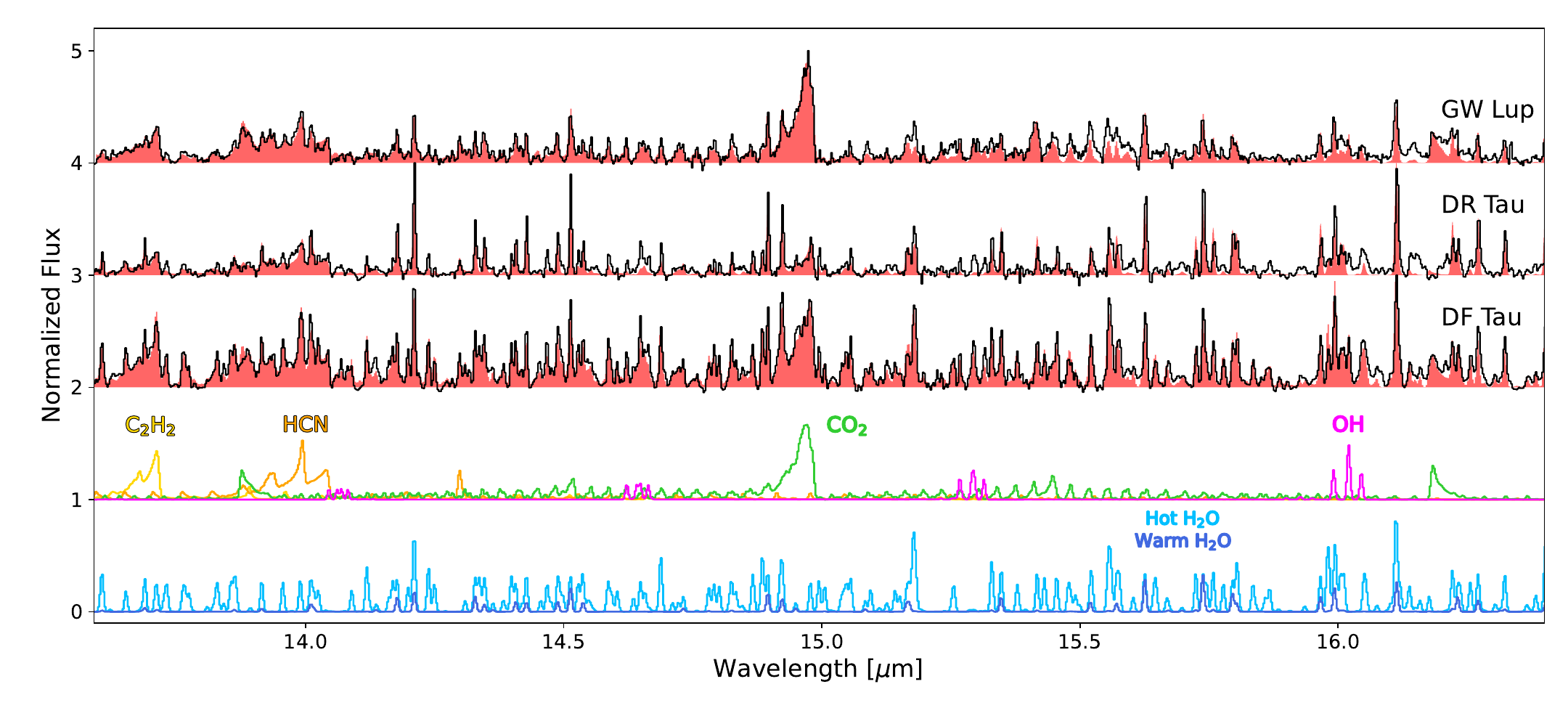}
\caption{Continuum-subtracted MIRI spectra (black) compared to a
  best-fitting total model in red, for the disks around GW Lup, a
  large dust disk with rings (top); DR Tau, a compact dust disk
  (second row); and DF Tau, a close binary (third row). The spectra
  are normalized to the peak emission in this wavelength range, 0.035
  Jy for GW Lup, 0.52 Jy for DR Tau, and 0.27 Jy for DF Tau (note 
  the much weaker lines of GW Lup). Model emission from C$_2$H$_2$
  (yellow), HCN (orange), CO$_2$ (green) and OH (pink) is presented in
  the fourth row for parameters that best fit the DF Tau disk
  emission. The bottom row shows two H$_2$O models in cyan (920 K) and
  blue (490 K). Data and models taken from
  \cite{Grant23,Temmink24CO,Temmink24H2O,Grant24}, see those papers
  for best fitting values. Figure by Sierra Grant.}
  \label{Fig1}
\end{figure}

Figure~\ref{Fig1} summarizes the 13-17 $\mu$m spectrum of three disks
around T Tauri stars: GW Lup, DR Tau and DF Tau
\citep{Grant23,Grant24,Temmink24H2O,Temmink24CO}. The bottom part shows the
simulated spectra of the various molecules that contribute to the
emission in this region. These so-called slab models assume that the
level populations of each molecule are in LTE at a single temperature
$T$ with the column density $N$ and emitting area $\Omega$ determining
the strength of the lines. Typical temperatures are 600--800 K for
C$_2$H$_2$ and HCN and 300--400 K for CO$_2$, with emission coming
from the inner $\sim$0.3 au. H$_2$O has many lines throughout the MIRI
wavelength range indicating a temperature gradient roughly as
$R^{-0.5}$ from $\sim 1200$ K down to $\sim$180 K over the 0.1 -- few
au range \citep{Gasman23,Temmink24H2O}, as also found in JDISCS 
\citep{Romero24}.

It is clear that there is significant diversity among disks: many T
Tauri disks are bright in H$_2$O lines (e.g., DR Tau, DF Tau) whereas
others show particularly strong CO$_2$ features (e.g., GW Lup),
including $^{13}$CO$_2$ \citep{Grant23} and even CO$^{18}$O (e.g., CX
Tau, Vlasblom et al.\ subm.). The detection of isotopologs indicates
that the main $^{12}$CO$_2$ band is optically thick and that the
CO$_2$/H$_2$O column density ratio is much higher than would be
suggested by the relative strengths of the lines. The CO$_2$
isotopologs are also fitted with colder temperatures suggesting that
they are located deeper in the disk \citep{Bosman22CO2}.

What causes this diversity?  DR Tau has a compact millimeter dust
disk, whereas DF Tau is a 9 au separation binary system with very
small truncated dust disks around both components. Thus, radial drift
of icy grains followed by water sublimation in the inner disk could
play a role for these disks (\S~\ref{sect:scenarios}). In contrast, GW
Lup has a large dust disk with dust rings that can trap the icy
pebbles at large radii well beyond the snowlines.

However, not all compact dust disks are bright in water lines. CX Tau
is an example of a drift-dominated disk with brighter CO$_2$ than
H$_2$O lines (Vlasblom et al., subm.). Thermochemical models suggest a
number of factors that can enhance one species over the other
\citep{Vlasblom24}. For example, a small cavity between the H$_2$O and
CO$_2$ snowlines can selectively remove H$_2$O. Also, temperature
plays a role (\S~\ref{sect:scenarios}): at low $T$, OH will react with
CO to form CO$_2$ rather than with H$_2$ to form H$_2$O. Finally, the
age of the disk can affect relative abundances \citep{Sellek24}:
CO$_2$-rich ices drift in later than H$_2$O-rich ices.

Another sign that the situation may be more complex than sketched in
\S~\ref{sect:scenarios} is the detection of water in the inner region
of the transitional disks. PDS 70 is a well-known disk with a large
cavity in which two young giant planets have been detected
\citep{Keppler18,Haffert19}. The prominent dust ring at 45 au is
thought to trap all the drifting icy pebbles and prevent them from
entering the small (few au) inner disk.  It therefore came as a
surprise when JWST clearly detected water at $T\approx 600$ K from the
inner disk \citep{Perotti23}. Similarly, emission of water and other
molecules is detected from the inner region of the transitional disk
SY Cha with a large 70 au cavity \citep{Schwarz24}. A systematic study
of several gapped T Tauri disks in the MINDS sample reveals emission
of water and other species from the inner disk in
all cases (Gasman et al., subm.). This suggests that gaps do not fully
block the delivery of oxygen- and carbon-containing gas and (small,
icy) dust to the inner disk: dust traps are clearly ``leaky'' even if
the dust and gas cavities are broad and deep.

Taken together, the data indicate that the gas in the inner disks of T
Tauri stars is generally oxygen rich, with C/O$<$1: H$_2$O is usually
the most abundant molecule, followed closely by CO$_2$. C$_2$H$_2$ and
HCN typically have about two orders of magnitude lower column
densities. CH$_4$ is generally not detected (Temmink, priv. comm.). OH
is never a significant reservoir of oxygen, although it is a unique
probe of the UV field and high-temperature chemistry
\citep{Tabone24}. Note that the various molecules are not necessarily
co-located, neither radially nor vertically in the disk, so any column
density ratios should be interpreted with caution.  More generally,
inner disk chemistry may be less strongly related to outer disk
substructures than previously thought.

\subsection{Disks around very low-mass stars and brown dwarfs}

The JWST spectra of disks around very low mass stars (VLMS, $<$0.3
M$_\odot$) and brown dwarfs ($<$0.08 M$_\odot$) are generally very
different from those around T Tauri stars. VLMS are of particular
interest since late M-type stars (later than M4) are the most common
stars in the Galaxy, and many rocky exoplanet studies focus on them,
with the Trappist-1 system being the most famous example. Early hints
for bright C$_2$H$_2$ emission from VLMS disks came from {\it Spitzer}
spectra \citep{Pascucci13}, but JWST reveals their rich hydrocarbon
emission in unprecedented detail.

The first JWST example was the J16053215-1933159 (J1605) disk which
shows two prominent broad bumps at 7.7 and 13.7 $\mu$m
\citep{Tabone23}. These can be ascribed to highly optically thick
C$_2$H$_2$ emission in its $\nu_4$+$\nu_5$ and $\nu_5$ bands, with
orders of magnitude higher column densities ($> 10^{20}$ cm$^{-2}$) at
525 K coming from the inner 0.033 au. Larger hydrocarbons such as
C$_4$H$_2$ and C$_6$H$_6$ (benzene) are also prominent and detected
for the first time.

Subsequent JWST studies have revealed other VLMS disk examples with
very rich hydrocarbon spectra, although not with such prominent
C$_2$H$_2$ bumps as J1605. Rather, the many $P$- and $R$-branch lines
of hydrocarbon molecules like CH$_4$ and C$_2$H$_4$ form an underlying
pseudo-continuum on which their $Q$-branches are detected
\citep{Arabhavi24,Kanwar24}. C$_2$H$_6$ is also detected in a disk for
the first time. Of the oxygen-bearing species, CO$_2$ is detected, but
CO only sporadically and H$_2$O is weak, if detected at all
\citep{Tabone23,Arabhavi24}. There appear to be a few exceptions: the
disk around the VLMS Sz 114 (0.17 M$_\odot$) shows prominent H$_2$O
lines resembling more that of T Tauri stars \citep{Xie23}.

Chemical modeling shows that such high abundances of hydrocarbon
molecules can only be achieved if C/O$>$1 in the emitting layers
\citep{Kanwar24model}. This means that either carbon is enhanced, or
oxygen is depleted, or both. Mechanisms for enhancing carbon include
the destruction of hydrocarbon grains at $\sim$500 K (a so-called
``soot'' line, \citealt{Li21}) or the drifting in of hydrocarbon-rich
ices that are subsequently eroded \citep{Tabone23,Mah24}. Options for
oxygen depletion focus on locking up most of the oxygen in H$_2$O ice
in a dust trap outside the water snowline. Quantification of these
models requires better constraints on the actual H$_2$O emission which
is hidden below the forest of hydrocarbon lines in VLMS disks
(Arabhavi et al., subm.). If C/O is indeed $>$1, this may have
consequences for the chemical composition of (terrestrial) planets
that are forming in the inner disk. As for Earth, the planets may be
poor in carbon if most of the solid carbon has been removed from the
disk in the form of carbon-rich gas.

\section{Conclusions}

The first two years of JWST data have revealed a large diversity in
the spectra of protoplanetary disks. It is clear that stellar mass,
the presence of dust traps in the inner and/or outer disk, radial
drift of icy pebbles, and the age and history of the system all play a
role in setting inner disk abundances where planets are formed. Much
larger samples are needed to reveal systematic trends, if any, and
linking them to exoplanet composition statistics. The community can
look forward to many more JWST disk spectra in the coming years. \\

\noindent {\it Acknowledgments.} The author is grateful to the entire
MINDS team and the MIRI consortium for many years of fruitful
collaborations. Special thanks to Sierra Grant for providing Figures
1 and 2. This work is supported by A-ERC grant 101019751 MOLDISK.

\end{document}